\documentclass[prx,aps,superscriptaddress,amssymb,amsmath,twocolumn]{revtex4-1}
\usepackage{amsmath}
\usepackage{amssymb}
\usepackage{amsthm}
\usepackage{amsfonts}
\usepackage{listings}
\usepackage{diagbox}
\usepackage{enumerate}
\usepackage{latexsym}
\usepackage{color}
\usepackage{xcolor}
\usepackage{bm}
\usepackage{hyperref}
\usepackage{graphicx}
\usepackage[FIGTOPCAP]{subfigure}
\hypersetup{
 pdfnewwindow=true, colorlinks=true,
 linkcolor=blue, anchorcolor=blue,
 citecolor=blue, filecolor=blue,
 menucolor=blue, urlcolor=blue}

\newcommand{\beginsupplement}{%
        \setcounter{table}{0}
        \renewcommand{\thetable}{S\arabic{table}}%
        \setcounter{figure}{0}
        \renewcommand{\thefigure}{S\arabic{figure}}%
     }

\newcommand{\webirvsp}{\href{https://github.com/zjwang11/irvsp}{{\ttfamily irvsp}}}

\def\ie{{\it i.e.},\ }

\def\MM{Ta$_2M_3$Te$_5$}

\def\Pd{Ta$_2$Pd$_3$Te$_5$}
\def\Ni{Ta$_2$Ni$_3$Te$_5$}

\input{epsf}

\begin{document}

\tolerance 10000

\newcommand{\vk}{{\bf k}}

\draft

\title{Quantum Spin Hall Effect in Ta$_2M_3$Te$_5$ ($M=$ Pd, Ni)}
\author{Zhaopeng Guo}
\thanks{These authors contributed equally to this work.}
\affiliation{Beijing National Laboratory for Condensed Matter Physics,
and Institute of Physics, Chinese Academy of Sciences, Beijing 100190, China}

\author{Dayu Yan}
\thanks{These authors contributed equally to this work.}
\affiliation{Beijing National Laboratory for Condensed Matter Physics,
 and Institute of Physics, Chinese Academy of Sciences, Beijing 100190, China}
\affiliation{University of Chinese Academy of Sciences, Beijing 100049, China}

\author{Haohao Sheng}
\affiliation{School of Materials and Physics, China University of Mining and Technology, Xuzhou 221116}

\author{Simin Nie}
\email{smnie@stanford.edu}
\affiliation{Department of Materials Science and Engineering, Stanford University, Stanford, California 94305, USA}

\author{Youguo Shi}
\email{ygshi@iphy.ac.cn}
\affiliation{Beijing National Laboratory for Condensed Matter Physics,
and Institute of Physics, Chinese Academy of Sciences, Beijing 100190, China}
\affiliation{Center of Materials Science and Optoelectronics Engineering, 
University of Chinese Academy of Sciences, Beijing 100049, China}

\author{Zhijun Wang}
\email{wzj@iphy.ac.cn}
\affiliation{Beijing National Laboratory for Condensed Matter Physics,
and Institute of Physics, Chinese Academy of Sciences, Beijing 100190, China}
\affiliation{University of Chinese Academy of Sciences, Beijing 100049, China}

\date{\today}

\begin{abstract}
Quantum spin Hall (QSH) effect with great promise for the potential application in spintronics and quantum computing has 
attracted extensive research interest from both theoretical and experimental researchers.
Here, we predict monolayer \Pd~can be a QSH insulator based on first-principles calculations.
The interlayer binding energy in the layered van der Waals compound Ta$_2$Pd$_3$Te$_5$ is 19.6 meV/\AA$^2$; thus, its monolayer/thin-film structures could be readily obtained by exfoliation.
The band inversion near the Fermi level ($E_F$) is an intrinsic characteristic, which happens between Ta-$5d$ and Pd-$4d$ orbitals without spin-orbit coupling (SOC). The SOC effect opens a global gap and  makes the system a QSH insulator.
With the $d$-$d$ band-inverted feature, the nontrivial topology in monolayer \Pd~is characterized by the time-reversal topological invariant $\mathbb Z_2=1$, which is computed by the one-dimensional (1D) Wilson loop method as implemented in our first-principles calculations. The helical edge modes are also obtained using surface Green's function method.
Our calculations show that the QSH state in \MM~($M=$ Pd, Ni) can be tuned by external strain.
These monolayers and thin films provide feasible platforms for realizing QSH effect as well as related devices.
\end{abstract}

\maketitle

\section{introduction}
A quantum spin Hall (QSH) insulator is a novel state of two-dimensional (2D) quantum matter that is interesting for both fundamental condensed matter physics and materials science. It has an insulating 2D bulk but conducting edge states. The backscattering channel for such edge states is prohibited by time-reversal symmetry (TRS).
The dissipationless conducting channels make it promising to realize low-power consumption electronics and spin-polarized current. In addition, on the interface of a conventional superconductor and a QSH insulator, superconducting proximity effect can lead to the appearance of topological superconductivity and Majorana fermions~\cite{fu2008superconducting}, which can be used for topological quantum computation~\cite{nayak2008non}. 
Due to these wide-ranging and fascinating applications, the QSH insulator has become one of the most hot and fruitful fields in condensed matter physics.

In despite of many proposals for QSH insulators~\cite{liu2011,chen2014,xu2013,zhang2018,ma2015robust,zhou2014large,luo2015room,weng2014transition,Qian1344}, 
the QSH state has been confirmed in very few systems, such as quantum well structures of HgTe/CdTe~\cite{Konig766} and InAs/GaSb~\cite{knez2011}, and monolayer 1T'-WTe$_2$~\cite{tang2017quantum,Wu2018}.
On the other hand, atomic monolayers of van der Waals materials (vdW monolayers) have intrigued various study of 2D materials recently, revealing a wide range of extraordinary properties and functionalities, such as 2D ferromagnetism~\cite{CrGeTe2017,FeGeTe2018a,FeGeTe2018b}, Ising superconductivity~\cite{NbSe2017,PdTe2020}, and unconventional superconductivity, etc~\cite{cao2018a,cao2018b}.
These atomic vdW monolayers are usually quite stable and easily exfoliated due to the weak vdW interaction between adjacent layers. 
Owing to their atomic thickness, these 2D layers can be readily tuned through chemical and mechanical techniques.
However, the vdW monolayers of the QSH state are interesting but very rare~\cite{Qian1344}.

In this paper, based on first-principles calculations, we find that 
the vdW monolayers \MM~($M=$ Pd, Ni) are potential candidates for realizing QSH effect. With experimental crystal parameters, our calculations show that monolayer \Pd~is a QSH insulator with a small band gap, while monolayer \Ni~is trivial. After applying uniaxial strain, monolayer \Ni~can be tuned into the QSH state. The interlayer binding energy in layered bulk Ta$_2$Pd$_3$Te$_5$ is 19.6 meV/\AA$^2$. Resultantly, monolayer \Pd~can be readily exfoliated from the bulk crystals.
The $d$-$d$ band inversion at $\Gamma$ already takes place without spin-orbit coupling (SOC).
Upon including SOC effect, monolayer \Pd~becomes a QSH insulator, which is characterized by the computed nonzero $\mathbb Z_2$ index with the 1D Wilson loop method~\cite{Wilsonloop2011}.
The existence of nontrivial helical edge states is confirmed by the open-boundary calculations. 
Our study also shows that external strain is a good means to engineer the QSH state in monolayers \MM.  
Our findings suggest that it is promising to realize QSH effect as well as related devices in these monolayers or thin films in the future.

\begin{figure*}[t]
\includegraphics[width=6.5in,height=2.5in]{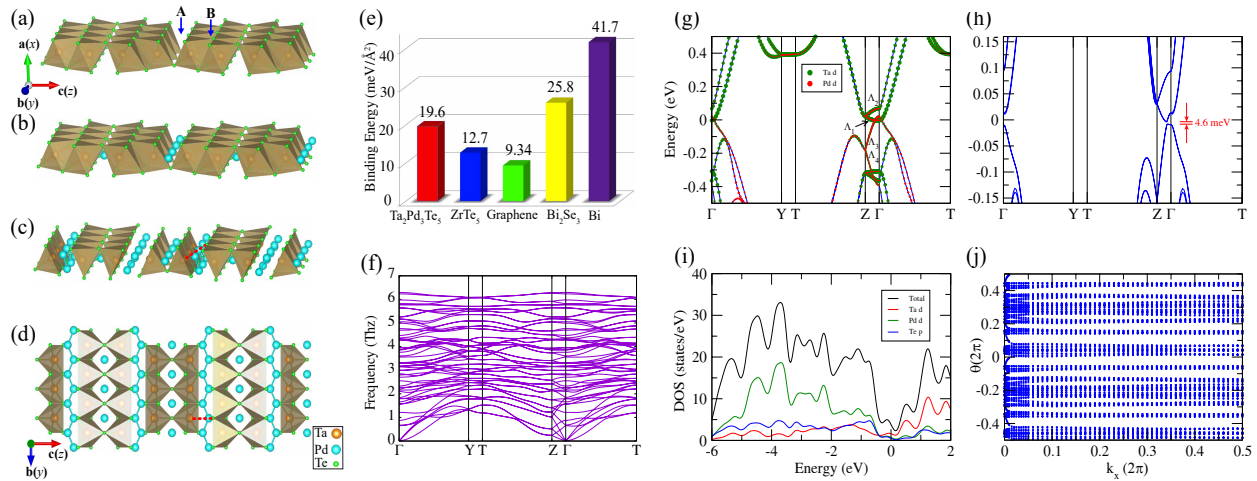}
      \caption{(color online).
       (a) Side view of Ta$_2X_5$ ($X=$~Se, Te) layer. Magenta and green balls represent Ta and $X$ atoms, respectively. The positions of A site and B site in Ta$_2X_5$ layer are marked by arrows.
       (b) Side view of Ta$_2$NiSe$_5$ layer. The A sites are filled by Ni atoms (cyan balls).
       (c) Side view of Ta$_2M_3$Te$_5$ ($M=$~Ni, Pd) layer.
       (d) Top view of monolayer Ta$_2$Pd$_3$Te$_5$. It has three symmetry operations: $\hat {g}_{x}\equiv\{M_x|0,1/2,1/2\}$, $\hat M_y$ and $\hat {C}_{2z}\equiv\{C_{2z}|0,1/2,1/2\}$.
       (e) The interlayer binding energies of Ta$_2$Pd$_3$Te$_5$ and other materials such as ZrTe$_5$, graphene, Bi$_2$Se$_3$, and Bi. (f) Phonon dispersion of monolayer Ta$_2$Pd$_3$Te$_5$.
       (g) The orbital-resolved PBE band structure of monolayer \Pd. The characters of Ta-$5d$ and Pd-$4d$ orbitals are marked by green and red dots, respectively. The irreps for four low-energy bands on $\Lambda$ line are denoted.
       (h) The PBE+SOC band structure of monolayer \Pd. A band gap of 4.6 meV is opened due to the SOC effect.
       (i) The calculated total DOS and partial DOS of monolayer \Pd. The black, red, green and blue lines represent the total DOS and partial DOS of Ta-$5d$, Pd-$4d$ and Te-$5p$ orbitals, respectively.
       (j) The evolution of WCCs for monolayer \Pd.}
\label{fig1}
\end{figure*}



\section{Calculation method and crystal structures}

The first-principles calculations were performed within the
framework of the projector augmented wave (PAW) method~\cite{bloch1994,kresse1999} implemented in
Vienna \emph{ab initio} simulation (VASP) package~\cite{kresse1996_1,kresse1996_2}. 
The Perdew-Burke-Ernzerhof (PBE) generalized gradient approximation exchange-correlations functional~\cite{PBE} was implemented in calculations.
The cut-off energy for plane wave expansion was 500 eV, and 1 $\times$ 16 $\times$ 4 $k$-point sampling grids were used in the self-consistent process (as the monolayer structure was placed in the $yz$-plane in our calculations).
SOC was taken into account within the second variational method self-consistently (PBE+SOC).
Single crystals of \Pd~were synthesized by self-flux method and the detailed crystallographic parameters were obtained by the single-crystal x-ray diffraction (XRD) study. 
The details of XRD study are shown in Section~\ref{sup:A} of the supplementary materials (SM).
To study the strain effect, the internal atomic positions have been fully relaxed until the residual forces on each atom were less than 0.01 eV/\AA~in the strained structures. 
The irreducible representations (irreps) were obtained by the program \webirvsp~\cite{gao2020}.
An $1\times 4\times 2$ supercell was built to calculate phonon dispersion using PHONOPY package \cite{togo2008}. The maximally localized Wannier
functions (MLWFs) for the Ta-$5d$, Pd-$4d$ and Te-$5p$ orbitals were constructed to calculate the edge states from the surface Green's function of the semi-infinite system based on the iterative method \cite{sancho1984quick,sancho1985highly, nicola2012}.
To investigate the interatomic bonding in the compounds, crystal orbital Hamilton populations (COHP) calculations were preformed using LOBSTER package~\cite{COHP1993,COHP2011,COHP2013,COHP2016-1,COHP2016-2,COHP2020}.

\begin{figure*}[htb]
	\includegraphics[width=7in]{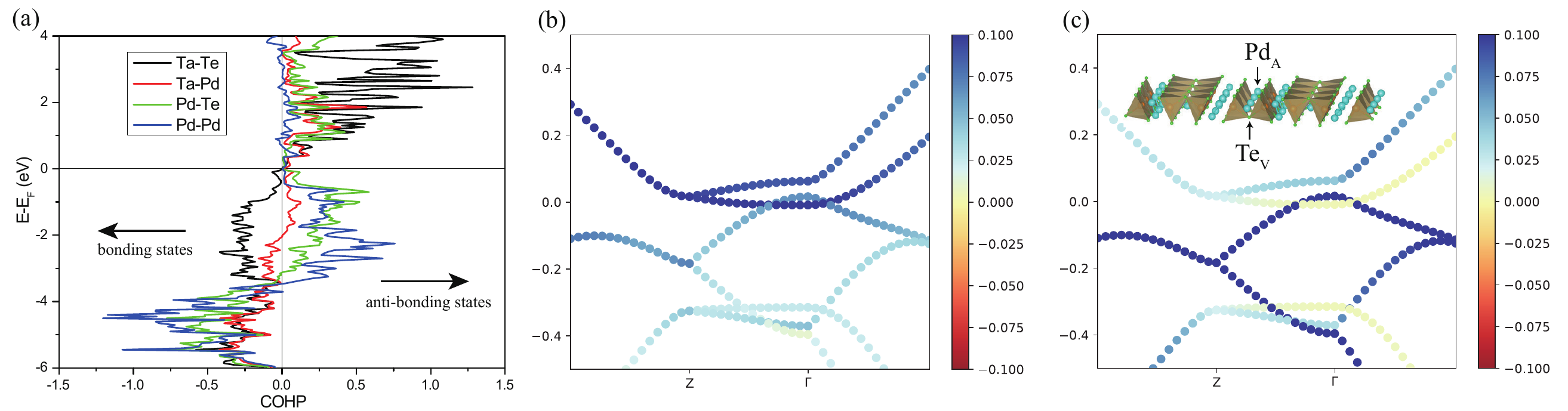}
	\centering
	\caption{(color online).
		(a) COHP diagram of monolayer \Pd.
		(b) The $k$-dependent COHP of Pd--Ta bonds.
		(c) The $k$-dependent COHP of Pd$_A$--Te$_V$ bonds. The insert shows the positions of Pd$_A$ and Te$_V$ atoms. Pd$_A$ and Te$_V$ represent the Pd atoms on the tetrahedral void A sites and the Te atoms on the vertexes of Te square pyramids, respectively.
	}\label{fig2}
\end{figure*}

The layered vdW compound Ta$_2$Pd$_3$Te$_5$ crystallizes in an orthorhombic layered structure with space group (SG) \#62 ($Pnma$ or $D_{2h}^{16}$) (See details in Section~\ref{sup:A} of the SM).
The bulk primitive unit cell contains two layers (\ie $yz$-plane), which are related by inversion symmetry. The adjacent layers are connected by vdW interactions.
To understand the crystal structure of monolayer \MM, we first construct an idealized polyhedral representation of the double chain of edge-sharing Ta$_2X_6$ ($X=$ Se, Te) octahedra, as shown in Fig.~\ref{fig1}(a). In this double chain structure, the double chain of Ta$_2X_6$ octahedra are condensed over their apical $X$ atoms to form a Ta$_2X_5$ layer. In the Ta$_2X_5$ layer, there are two types of tetrahedral voids (\ie A site and B site) which are capable of accepting ``interstitial" metal atoms. 
Then, based on Ta$_2X_5$ layer, Ta$_2$NiSe$_5$ layer [Fig.~\ref{fig1}(b)] can be understood by filling one A site with nickel, while \MM~layer [Fig.~\ref{fig1}(c)] can be described with one A site and two B sites filled with $M$ atoms.
Thus, in \MM~layer, the Ta atoms are actually coordinated to five Te neighbors in an approximate square pyramid at distances between 2.7 and 2.8 \AA, with an additional sixth Te atom at a nonbonding [red dashed lines in Figs.~\ref{fig1}(c) and \ref{fig1}(d)] Ta--Te distance of 3.9 \AA~due to the insertion of the Pd chain at B sites.
Fig.~\ref{fig1}(d) shows a top view of the monolayer \Pd~(\ie the projection along [100]). 
Whereas the metal atoms in practically all known four-coordinate palladium complexes are square-planar coordinated as a result of the strong ligand-field splitting~\cite{vanquickenborne1981ligand}, the palladium in the \Pd~is coordinated tetrahedrally to the chalcogenides, which is rather unusual. The metal-metal bonding interactions between Ta and $M$ metals are of crucial rule for the electronic stability of these structures~\cite{tremel1993tetrahedral}, which are confirmed by our COHP calculations in Fig.~\ref{fig2}(a).

\section{results and discussions}

The bulk \MM~is of orthorhombic structure with $Pnma$ (No. 62). The symmetry classification for SG 62 is $\mathbb Z_2\times \mathbb Z_2\times \mathbb Z_2\times \mathbb Z_4$~\cite{song2018si}, which are computed to be (0000) for both \Pd~and \Ni. However, their mirror Chern numbers in $k_y=0$ plane are computed to be nonzero. These results indicate that the bulk \MM~are topological crystalline insulators with trivial symmetry indicators (see more details in Section~\ref{sup:B} of the SM).
The interlayer binding energy of Ta$_2$Pd$_3$Te$_5$ is computed to be about 19.6 meV/$\text{\AA}^2$, which is smaller than of 1T'-WTe$_2$ (28 meV/$\text{\AA}^2$) and compatible with other layered materials~\cite{weng2014transition,nie2019topological}, as shown in Fig.~\ref{fig1}(e). Given the weak interlayer coupling strength, monolayers and thin films of \Pd~can be exfoliated easily from the bulk materials. To confirm the stability of the monolayer, its phonon spectrum has been obtained in Fig.~\ref{fig1}(f). No imaginary frequency mode in the phonon spectrum indicates that the monolayer is dynamically stable.
We focus on the results of monolayers \MM~in the main text and put the results of the 3D bulks in the SM.

To begin with, the PBE band structure of monolayer Ta$_2$Pd$_3$Te$_5$ is presented in Fig. \ref{fig1}(g) with different orbital weights. 
Along the Z--$\Gamma$ line (\ie $\Lambda$ line), the low-energy bands near $E_F$ are two conduction bands ($\Lambda_2$ and $\Lambda_1$ irreps) derived from Ta-$5d$ orbitals, and two valence bands ($\Lambda_3$ and $\Lambda_4$ irreps) originating from Pd-$4d$ orbitals (neglecting the contributions from Te atoms for simplicity).
The irreps are labeled by the single point group $C_{2v}$ on $\Lambda$ line.
The calculated density of states (DOS) are shown in Fig.~\ref{fig1}(i) together with the partial DOS for Ta-$5d$, Pd-$4d$ and Te-$5p$ orbitals.
The results suggest that Te-$4p$ and Pd-$4d$ states are occupied below $E_F$, while the states above $E_F$ are mainly dominated by Ta-$5d$ orbitals. It's noted that the Pd-$4d$ and Ta-$5d$ states have strong hybridization around $E_F$.
The band crossing between $\Lambda_3$ and $\Lambda_1$ bands, which have different $\hat{g}_{x}$ eigenvalues, results in a $\hat{g}_{x}$-protected nodal line in the 2D Brillouin zone (BZ). Thus, there are real crossing points along $\Gamma$--Y and $\Gamma$--T in Fig.~\ref{fig1}(g).

Upon including SOC, the nodal line opens a band gap ($\sim$ 4.6 meV) and the system becomes an insulator, as shown in Fig.~\ref{fig1}(h).
Due to the lack of inversion symmetry, the $\mathbb Z_2$ index in monolayer \Pd~is computed through the 1D Wilson loop method. The computed Wannier charge centers (WCCs) in Fig.~\ref{fig1}(j) show a zig-zag pattern in half of the BZ, indicating that monolayer \Pd~is a QSH insulator.

To further understand the band inversion mechanism of monolayer \Pd, we calculated 
the COHP of Ta--Te, Ta--Pd, Pd--Pd and Pd--Te chemical bonds, as shown in Fig.~\ref{fig2}(a).
The positive and negative values of COHP indicate anti-bonding and bonding states, respectively.
As expected, the bonding (anti-bonding) states of Ta--Te bands are below (above) $E_F$. 
Additionally, the anti-bonding states of Pd--Ta and Pd--Te bonds also make many contributions below $E_F$.
By calculating the $k$-dependent COHP for the low-energy bands in Figs.~\ref{fig2}(b,c), we find that the anti-bonding states of Ta--Te bonds contribute to the two conduction bands, while the two valence bands are mainly from the anti-bonding states of Pd$_A$--Te$_V$ bonds (the Ta--Te bonding states are neglected for simplicity).
Thus, we conclude that the band inversion is closely related to the Pd$_A$--Te$_V$ anti-bonding states. 
To further verify this conclusion, we have investigated the evolution of the band structures by changing the length of the Pd$_A$--Te$_V$ bond (\ie simply moving Pd$_A$ in $x$-direction). The results are shown in Fig.~\ref{fig:s5}. As Pd$_A$ atom moves away from Te$_V$ atom, the Pd$_A$--Te$_V$ bond is weakened due to the increase of the bond length. Thus, the anti-bonding states (valence bands) move downwards energetically, which would remove the band inversion.  
We find that the band inversion could be an intrinsic characteristic of the A-site filled structure and already takes place without including SOC in the first-principles calculations.
Our PBE calculations show that monolayers \Pd~and \Ni~are near the phase boundary between a trivial insulator and a QSH insulator.

\begin{figure}[tb]
	\includegraphics[angle=0,width=3.5in]{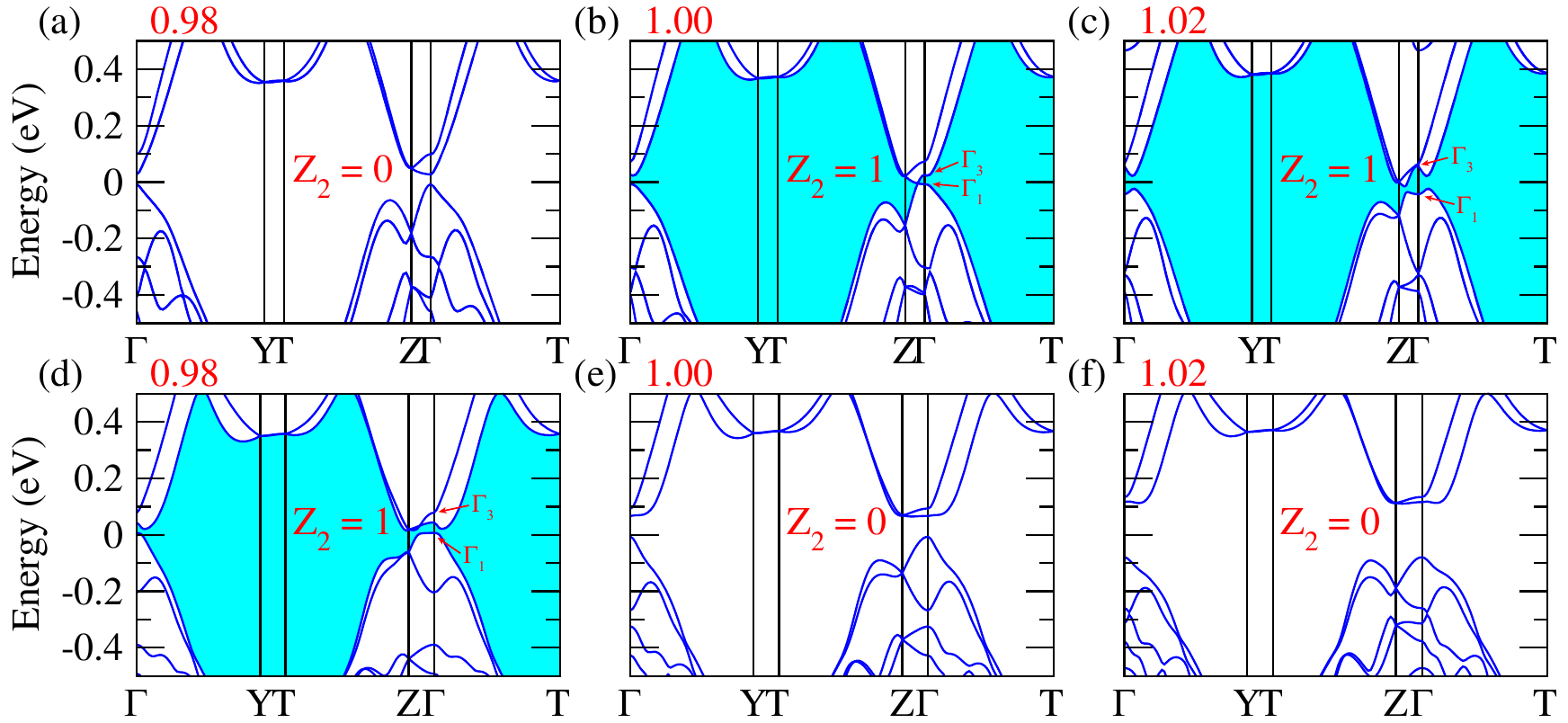}
	\caption{(color online).
		(a-c) The band structures of monolayer \Pd~with uniaxial strain along $z$-direction: $\eta c_0$ with $\eta= 0.98,~1,~1.02$, respectively. Since there is only one doubly-valued irrep at $\Gamma$ ( the double point group $C_{2v}$), we prefer to label the bands of these band structure plots by the irreps without SOC. (d-f) The band structures of monolayer \Ni~with uniaxial strain along $y$-direction: $\eta b_0$ with $\eta= 0.98,~1,~1.02$, respectively.}
	\label{fig3}
\end{figure}

When these vdW monolayers are grown on different substrates, the strain can be easily applied on monolayer samples. Thus, we investigate the strain effect on the systems. Fig.~\ref{fig3} shows the PBE+SOC band structures of monolayers \Pd~and \Ni~under different uniaxial strains. 
The results show that the QSH state can be tuned by external strain. 
With uniaxial strain along $z$-direction ($c=1.02c_0$), the band inversion (defined as $E_g\equiv |E_{\Gamma_1}-E_{\Gamma_3}|$) at $\Gamma$ becomes larger, as shown in Fig.~\ref{fig3}(c). Although monolayer \Ni~is a trivial insulator without strain, it would become a QSH insulator after applying a small compressive strain along $y$-direction ($b=0.98b_0$) in Fig.~\ref{fig3}(d) ($b_0$ and $c_0$ are the lattice constants obtained by our XRD study in Section~\ref{sup:A} of the SM). Therefore, we find that compressive strain along $y$-direction (and tensile strain along $z$-direction) benefits the $d$-$d$ band inversion at $\Gamma$ and the emergence of the resultant QSH state.

\begin{figure}[tb]
	\includegraphics[angle=0,width=3.3in]{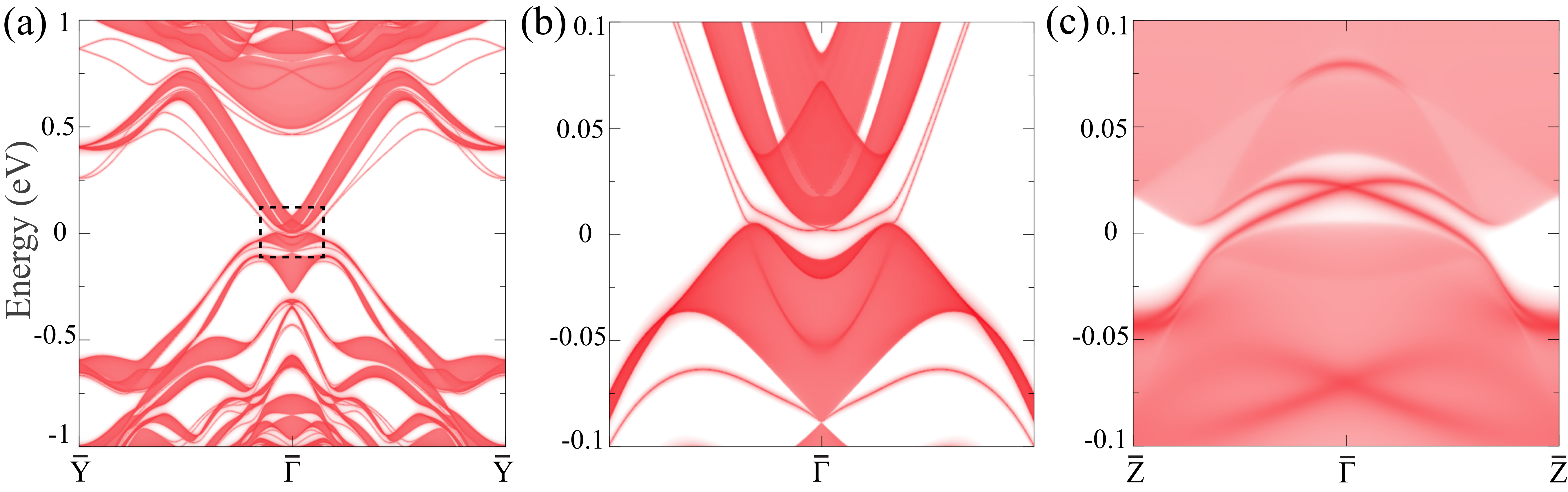}
	\caption{(color online).
		The edge states of monolayer \Pd~under 2\% tensile strain. 
		(a) The edge states along the $y$-direction.
		(b) The zoom-in plot of the dashed box in (a).
		(c) The edge states along the $z$-direction.}
	\label{fig4}
\end{figure}

In view of the fact that the hallmark of topological nontrivial property for the QSH state is the existence of topologically nontrivial edge states.
The tight-binding Hamiltonians of semi-infinite samples have been constructed by the MLWFs for Ta-$5d$, Pd-$4d$ and Te-$5p$ orbitals, which were extracted from the first-principles calculations for monolayer \Pd~under 2\% tensile strain along $z$-direction. 
The imaginary parts of the surface Green's functions show the spectrum of the edge states. The $y$-directed and $z$-directed edge spectra are shown in Figs.~\ref{fig4}(a) and \ref{fig4}(c), respectively. In the bulk gap (Figs.~\ref{fig4}(b-c)), there are two helical edge states connecting the bulk valence continuum to the conduction continuum, which are consistent with the nontrivial $\mathbb Z_2$ topology of the system.

As the energy band gap can be underestimated in the PBE calculations and more complicated details should be considered in the future experiments, we have the following discussions on the QSH state in \MM. 
First, the band structures are sensitive to the external strain, which provides a good means to manipulate the QSH state. Second, once considering few-layer thin films, the strength of band inversion is enhanced (see details in Section~\ref{sup:E} of the SM), which is helpful to realize the QSH state. Third, atomic thin monolayer offers a facile mechanism to control the electronic properties by electric field in vdW devices.
Last, considering layered vdW heterostructures with ferromagnetic 2D layers, they can be used to design novel topological states in the future, such as quantum anomalous Hall effect.
In addition, we have undertaken scanning tunneling spectroscopy measurements. Although the details will be published separately, we note that the edge states are directly observed at the step edges of the \Pd~crystals, which agree well with the QSH state of monolayer \Pd.

\section{CONCLUSION}
In summary, we performed first-principles calculations for the electronic structures and topological property of monolayers \MM. The band inversion in \Pd~has already happened without SOC, resulting in a nodal line surrounding $\Gamma$, which is protected by the glide mirror $\hat{g}_{x}$. Upon including SOC, it becomes a QSH insulator with a global band gap. The $\mathbb Z_2$ invariant is computed to be 1.
The helical edge states are obtained accordingly.
In addition, the study of the strain effect shows that the band structures and band topology in monolayers \MM~can be easily tuned by uniaxial strain.
The monolayer \Ni~can also be a QSH insulator as well under 2\% uniaxial strain along $y$-direction.
Considering the stability and transferability of the atomic layers, the monolayers and thin films of \MM~provide feasible platforms to study QSH effect as well as related devices.

\begin{acknowledgments}
This work was supported by the National Natural Science Foundation of China (Grants No. 11974395, U2032204), the Strategic Priority Research Program of Chinese Academy of Sciences (Grant No. XDB33000000), and the CAS Pioneer Hundred Talents Program.
Y.G. Shi acknowledges the support from Chinese National Key Research and Development Program (No. 2017YFA0302901, 2016YFA0300604), the K. C. Wong Education Foundation (GJTD-2018-01) and Beijing Natural Science Foundation (Z180008).
\end{acknowledgments}

\bibliography{refs}


\clearpage

\begin{widetext}

\beginsupplement{}
\setcounter{section}{0}
\section*{APPENDIX}

\subsection{Crystal growth and structural measurement}
\label{sup:A}
Single crystals of \Pd~were synthesized by self-flux method. 
Starting materials of Ta (powder, 99.999\%), Pd
(bar, 99.9999\%) and Te (lump, 99.9999\%) were mixed in an Ar-filled glove box at a molar radio of Ta : Pd : Te = 2 : 4.5 : 7.5.
The mixture was placed in an alumina crucible, which was then sealed in an evacuated quartz tube.
The tube was heated to 950 $^{\circ}$C over 10 h and dwelt for 2 days.
Then, the tube was slowly cooled down to 800 $^{\circ}$C at
a rate of 0.5 $^{\circ}$C/h.
Finally, the extra flux was removed by centrifuging at 800 $^{\circ}$C.
After centrifuging, the black and
shiny single crystals of \Pd~can be picked out from the remnants in the crucible.

To investigate the crystalline structure, single-crystal XRD was carried out on Bruker D8 Venture
diffractometer at 273 K using Mo K$\alpha$ radiation ($\lambda$ = 0.71073 \AA).
The crystalline structure was refined by full-matrix
least-squares method on $F^2$ by using the SHELXL-2018/3 program.

The single-crystal XRD study revealed that \Pd~is of orthorhombic structure with SG $Pnma$ (No. 62).
The lattice parameters of \Pd~are $a_0$ = 13.9531(6) \AA, $b_0$ = 3.7038(2) \AA~and $c_0$ = 18.5991(8) \AA.
The
detailed crystallographic parameters are summarized in Table \ref{table:s1}.
Fig.~\ref{fig:s1} shows the XRD pattern of a flat surface of
\Pd~single crystal, where only ($h$00) peaks are detected.
A photograph of a typical \Pd~crystal is shown
in the inset of Fig.~\ref{fig:s1}, and the back square of $1\times 1$ mm indicates the size of the crystal.
Fig.~\ref{fig:s2} shows the crystal structure of \Pd~in our measurements.


\begin{table}[htbp]
	\setlength{\tabcolsep}{1.5mm}
		\caption{Atomic coordinates and equivalent isotropic thermal parameters of \Pd.}
		\begin{tabular}{c c c c c c c}%
		\hline
		\hline
		Site  & $Wyckoff$ & $x$ & $y$ & $z$ & $Occup.$ & $U_{eq}$\\
		\hline 
		Ta1 & 4$c$ & 0.24318 & 0.75000 & 0.33301 & 1.000 & 0.010\\
		Ta2 & 4$c$ & 0.25092 & 0.25000 & 0.08293 & 1.000 & 0.010\\
		Te1 & 4$c$ & 0.13221 & 0.75000 & 0.45953 & 1.000 & 0.013\\
		Te2 & 4$c$ & 0.40024 & 0.25000 & 0.55902 & 1.000 & 0.012\\
		Te3 & 4$c$ & 0.38685 & -0.25000 & 0.13913 & 1.000 & 0.012\\
		Te4 & 4$c$ & 0.39487 & 0.25000 & 0.35369 & 1.000 & 0.012\\
		Te5 & 4$c$ & 0.39464 & -0.25000 & 0.77984 & 1.000 & 0.012\\
		Pd1 & 4$c$ & 0.32146 & 0.25000 & 0.68820 & 1.000 & 0.014\\
		Pd2 & 4$c$ & 0.26901 & 0.25000 & 0.45802 & 1.000 & 0.013\\
		Pd3 & 4$c$ & 0.30827 & 0.25000 & 0.22723 & 1.000 & 0.013\\
		\hline
		\hline
		\end{tabular}
		\label{table:s1}
\end{table}

\begin{figure}[htbp]
	\includegraphics[width=3.5in]{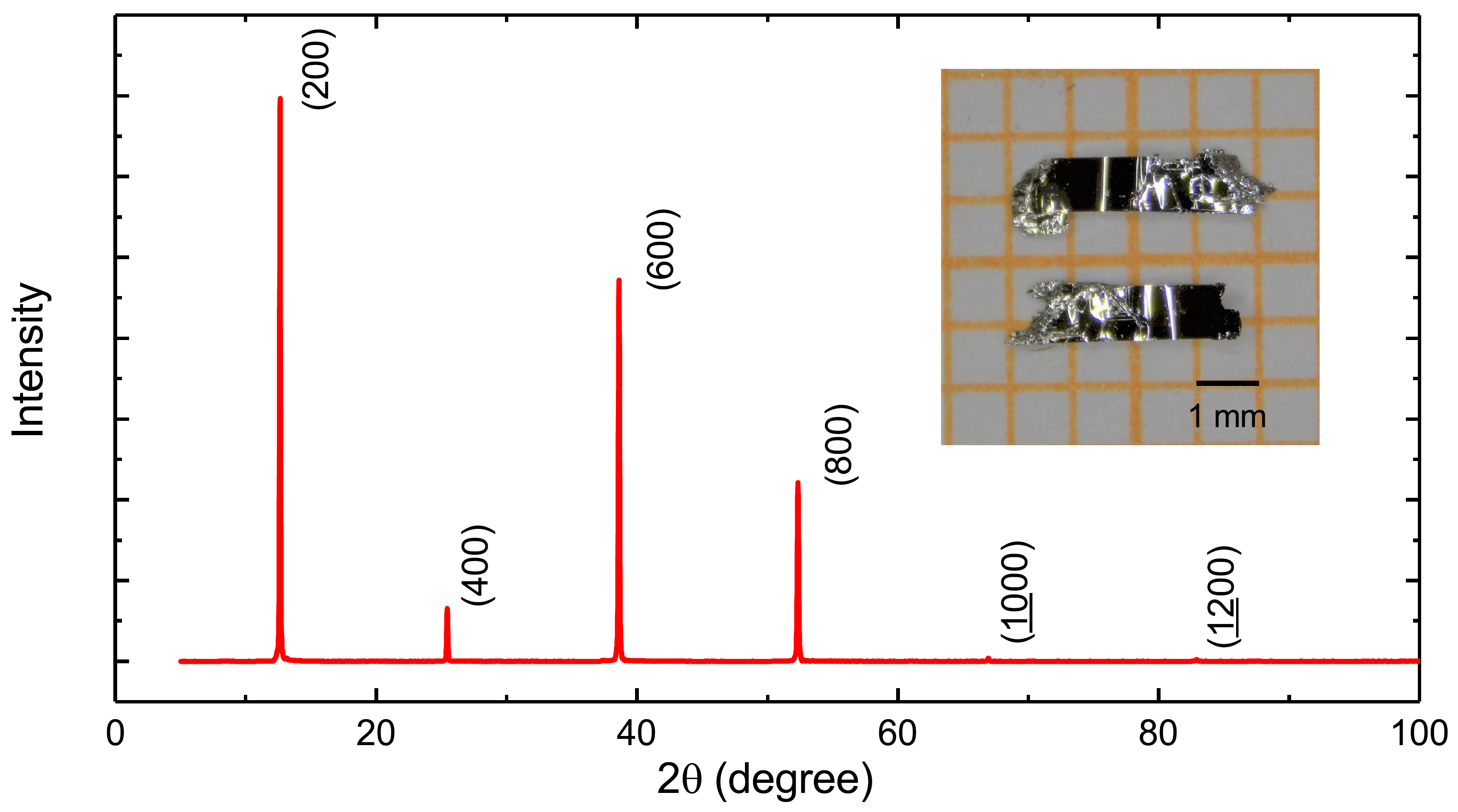}
	\caption{(color online).
		The XRD pattern of a flat surface of \Pd~single crystal. The inset shows a photograph of a typical
		\Pd~single crystal.
	}\label{fig:s1}
\end{figure}

\begin{figure}[htbp]
	\includegraphics[width=3.5in]{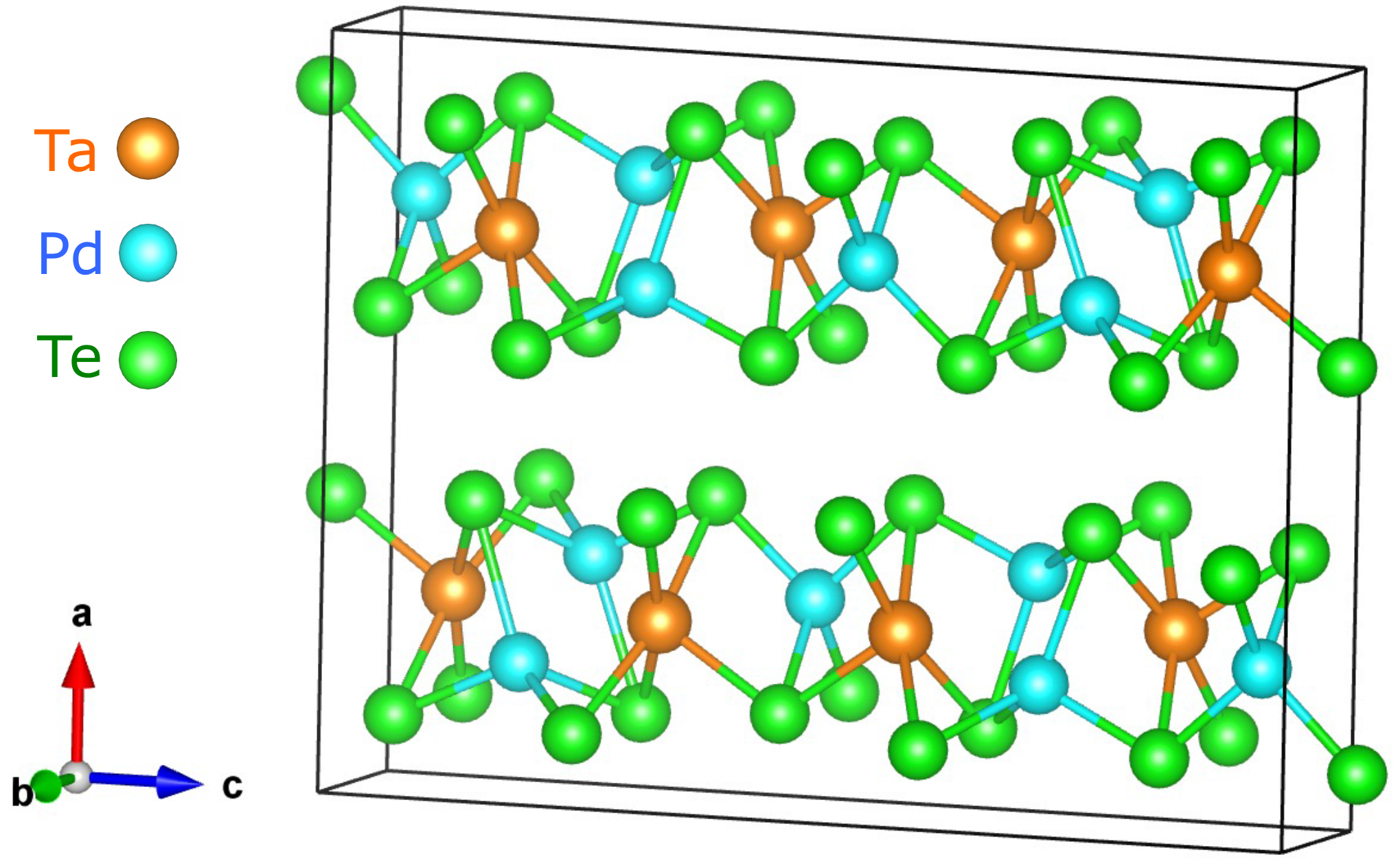}
	\caption{(color online).
     The crystalline structure of \Pd~obtained by our single-crystal XRD study.
	}\label{fig:s2}
\end{figure}
~\\
\subsection{Electronic band structures and band topology of layered vdW compounds \MM~}
\label{sup:B}
The bulk \MM~is a layered vdW material (\ie \MM~layers stacking in the [100] direction)~\cite{tremel1991isolierte}. The PBE+SOC band structures for \Ni~and \Pd~are given in Fig.~\ref{fig:s3}. As each unit cell consists two layers, four symmetry indicates ($\mathbb Z_2 \times\mathbb Z_2\times\mathbb Z_2\times\mathbb Z_4$) are computed to be 0. These symmetry indicates are also zero for the bulk \Ni. However, the 3D bulk crystal has a mirror symmetry $\hat M_y$. So the eigen-wavefunctions can be classified into two different subspaces by the $\hat M_y$ eigenvalues (\ie $\pm i$), where the Chern numbers ($C_{\pm i}$) are well defined.
Thus, the mirror Chern number $C_M$ is defined as $C_M\equiv\frac{C_i-C_{-i}}{2}$ ($C_M=C_i=-C_{-i}$) in the time-reversal invariant systems.
It can be calculated by the 1D Wilson loop method. 
The calculated WCCs for the $k_y=0$ plane shown in Figs.~\ref{fig:s4}(a,b) suggest $C_{M,0}=2$ in \Ni~and  $C_{M,0}=4$ in \Pd. The mirror Chern number for the $k_y=\pi/b$ is zero ($C_{M,\pi}=0$), whose WCCs are not presented.
The nontrivial mirror Chern number in \Ni~indicates the band inverted feature in its 3D crystal.

\begin{figure}[h]
	\includegraphics[width=4.5in]{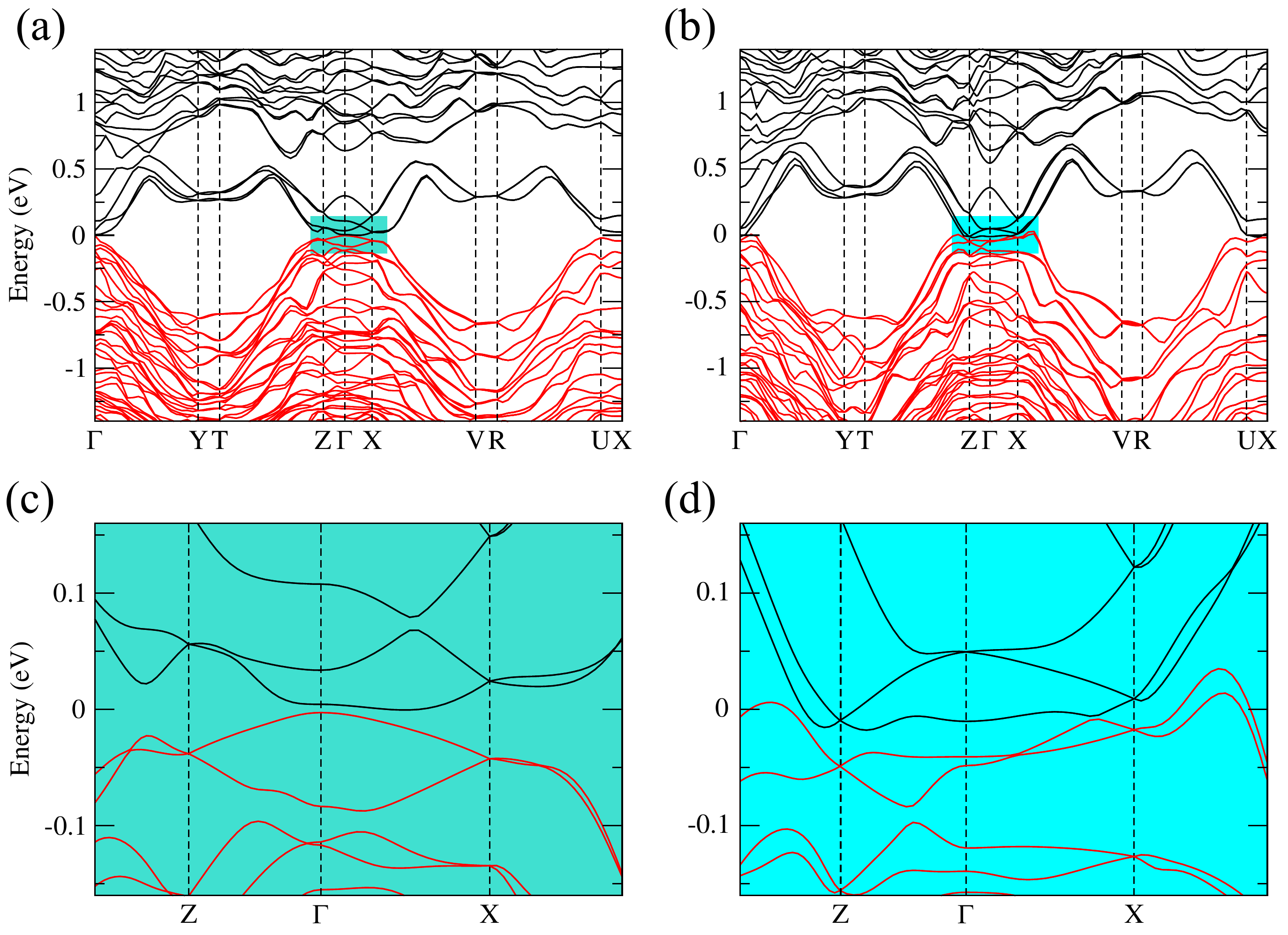}
	\caption{(color online).
    The PBE+SOC band structures of bulk \MM. 
    (a,c) The PBE+SOC band structures of bulk \Ni~and zoom-in plot of the shadowed area.
    (b,d) The PBE+SOC band structures of bulk \Pd~and zoom-in plot.
	}\label{fig:s3}
\end{figure}

\begin{figure}[htbp]
	\includegraphics[width=6.5in]{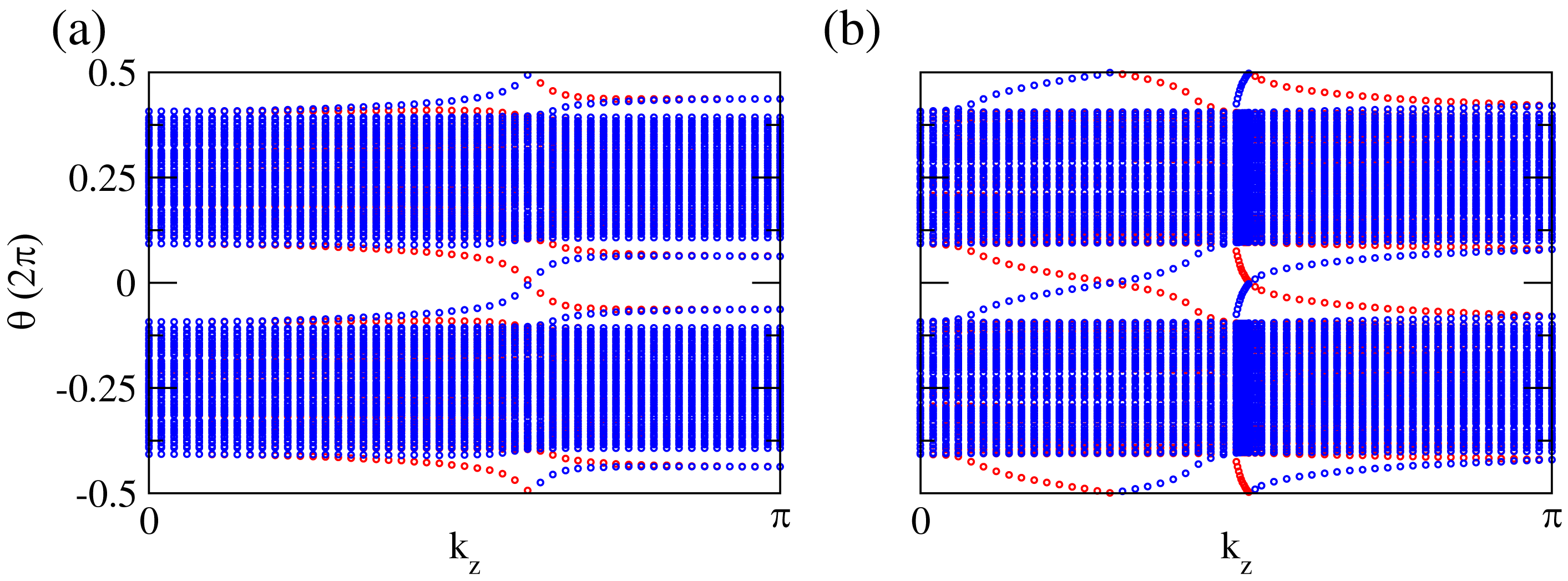}
	\caption{(color online).
     The evolution of WCCs for bulk (a) \Ni~and (b) \Pd~at $k_y=0$ plane. The blue and red circles mark the $+i$ and $-i$ subspaces of the $\hat M_y$ eigenvalues, respectively.
	}\label{fig:s4}
\end{figure}

\subsection{The evolution of band structures when moving Pd$_A$ atoms}
\label{sup:C}
The experimental length of the Pd$_A$-Te$_V$ bond is $d_0=2.66$ \AA. By simply shifting the Pd$_A$ atom in the $x$-direction by $\delta$, one can change the length of Pd$_A$-Te$_V$ bond.

\begin{figure}[htbp]
	\includegraphics[width=7in]{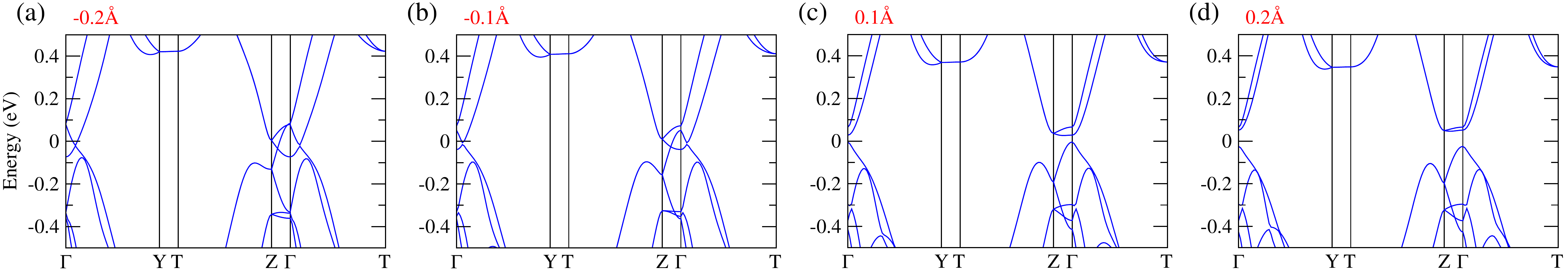}
	\caption{(color online).
		The PBE band structures of monolayer \Pd~with Pd$_A$ atoms moving along $x$-direction (perpendicular to \Pd~monolayer). The numbers of left upper corner mark the distances of Pd$_A$ atoms move away from \Pd~monolayer. 
	}\label{fig:s5}
\end{figure}

\subsection{The evolution of band structures with uniaxial strains}
\label{sup:D}

Under uniaxial strains, the band inversion of monolayer can be tuned easily. The related band structures of monolayers \Pd~(Fig.~\ref{fig:s6}) and \Ni~(Fig.~\ref{fig:s7}) are calculated.

\begin{figure}[htbp]
	\includegraphics[width=3.5in]{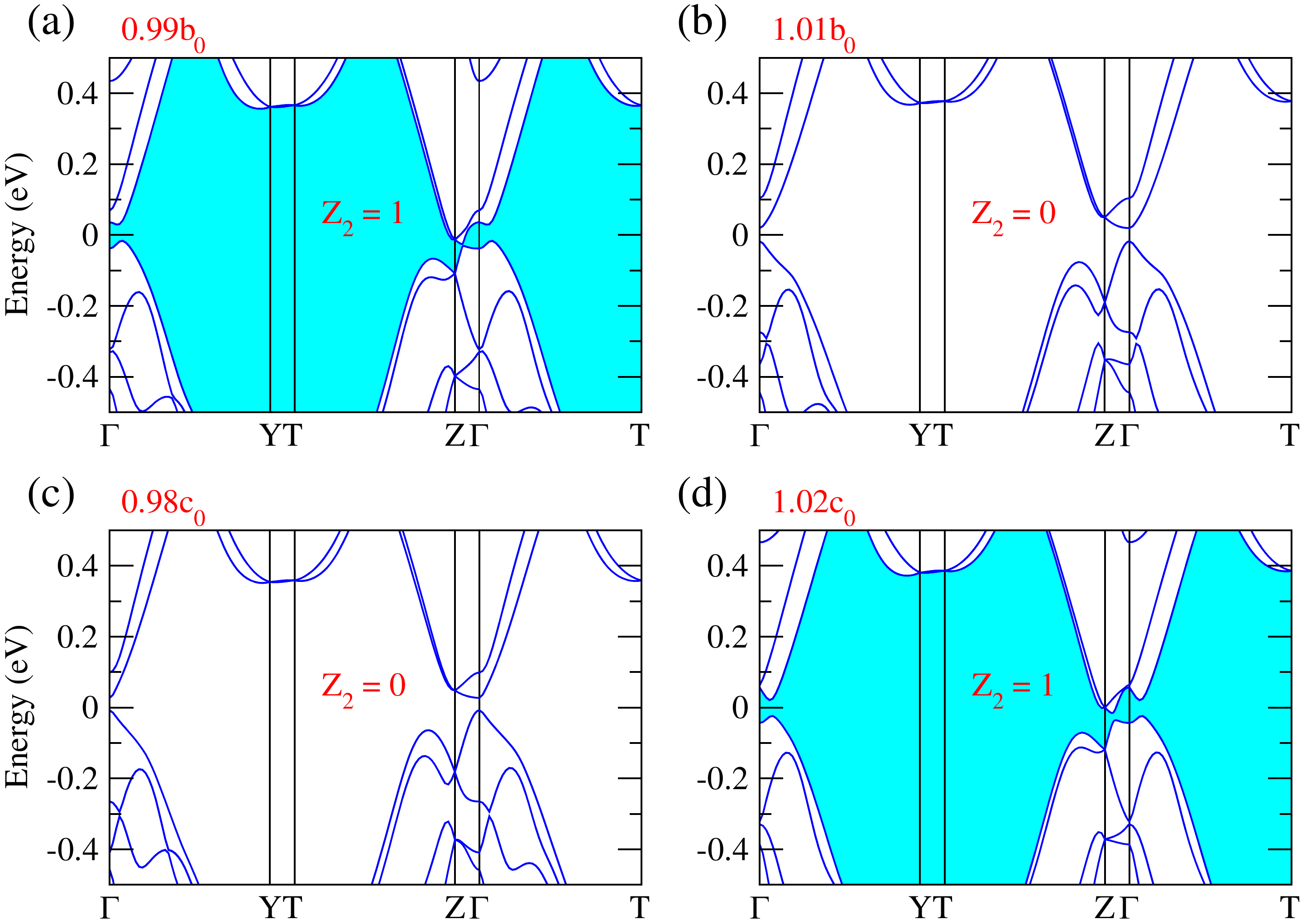}
	\caption{(color online).
		The PBE+SOC band structures of monolayer \Pd~under uniaxial strain along (a,b) $y$-direction and (c,d) $z$-direction, respectively.
	}\label{fig:s6}
\end{figure}

\begin{figure}[htbp]
	\includegraphics[width=3.5in]{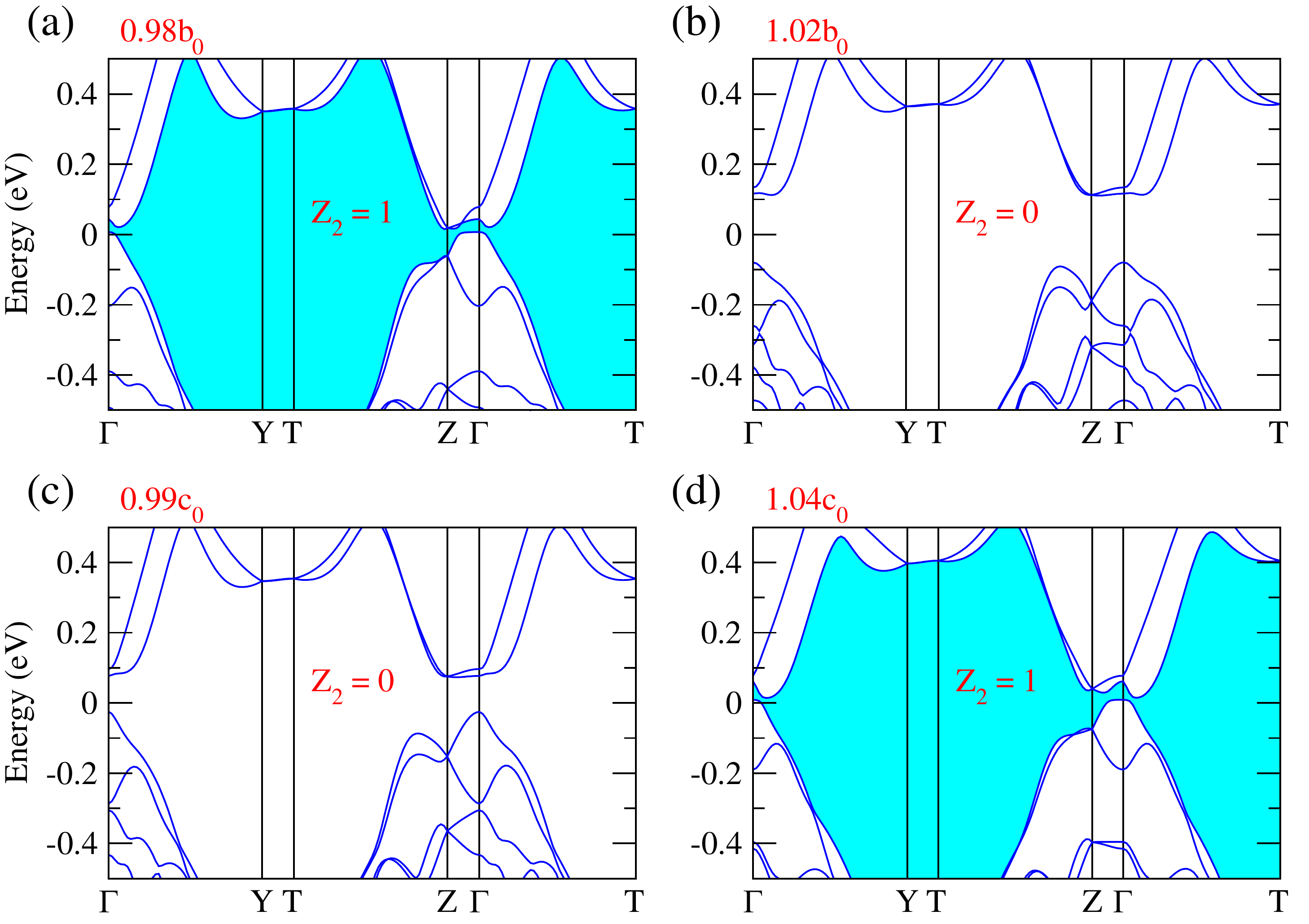}
	\caption{(color online).
		The PBE+SOC band structures of monolayer \Ni~under uniaxial strain along (a,b) $y$-direction and (c,d) $z$-direction, respectively.
	}\label{fig:s7}
\end{figure}

\clearpage

\subsection{The band structures for few layers}
\label{sup:E}
In the thin films, depending on the number of band inversions associated with the number of the layers, the few-layer systems show cross over between trivial and nontrivial 2D insulators according to film thickness.

\begin{figure}[htbp]
\includegraphics[width=6.8in]{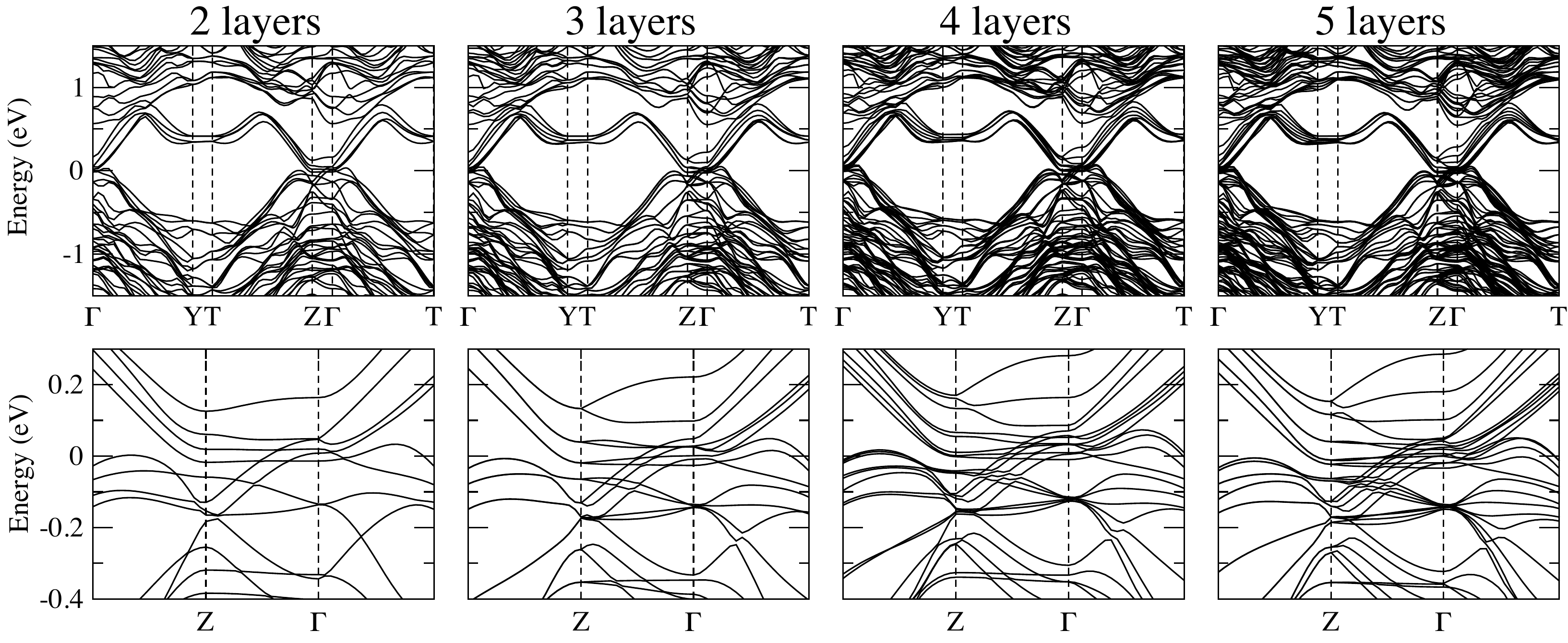}
   \caption{(color online).
   The PBE band structures for the slabs with different layers of \Pd. As the thickness of the slab increases, the strength of the band inversion is enhanced.
}\label{fig:s9}
\end{figure}

\newpage
\end{widetext}
\end{document}